*Molecular Self-Assembly of Jointed Molecules on a Metallic Substrate: From Single Molecule to Monolayer*[**]

**Tomaso Zambelli, Sylvain Goudeau, Jérôme Lagoute**[#]**, André Gourdon, Xavier Bouju, and Sébastien Gauthier**[*]

[*]   Dr. T. Zambelli, Dr. S. Goudeau, Dr. J. Lagoute, Dr. A. Gourdon,
      Dr. X. Bouju, Dr. S. Gauthier
      Groupe Nanosciences, CEMES-CNRS
      29 rue Jeanne Marvig, F-31055 Toulouse, France
      Fax: (+33) 5 62257999
      E-mail: zambelli@cemes.fr

[#]   Present address
      Paul Drude Institut für Festkörperelektronik
      Hausvogteiplatz 5-7
      D-10117 Berlin, Germany

[**]   This work was partly funded by the NMP Programme of the European Community Sixth Framework Programme for RTD activities, under the STREP project NANOMAN (*Control, manipulation, and manufacture on the 1-10 nm scale using localized forces and excitations*), contract No. NMP4-CT-2003-505660.

Because of its promising contribution to the bottom-up approach for nanofabrication of complex molecular architectures, self-organization is widely studied nowadays.[1,2] Numerous studies have tackled supramolecular chirality[3-6] or low-dimensional molecular nanostructures[7-14] using in most cases *small and rigid* molecules adsorbed on metallic substrates. In this situation, self-assembled structures can be understood in relative simple terms considering molecule-molecule versus molecule-substrate interactions. In contrast, the case of *large* and *three-dimensional molecules* which can adopt different adsorption conformations is more complex.[15,16] Here, we investigate the self-assembly of *V*-Landers[17] (*VL*) molecules ($C_{108}H_{104}$) (Fig. 1a and b) on Cu(100) by STM at room temperature (RT) under ultrahigh vacuum (UHV). This molecule is constituted of a central poly-aromatic board linked by sigma bonds to four 3,5-di-*tert*-butylphenyl (DTB) "legs".[18] It was demonstrated previously that the (physisorbed) adsorption conformations of a single molecule can be described in the framework of a simplified mechanical model based on the observation that the board and the legs behave as mechanically frozen blocks linked by flexible bonds.[19] The position of each leg is then described by only two angles. These "flexure hinges" were shown to play an essential role in the adaptability of the molecule to the substrate[19] and in its diffusion properties.[20] These degrees of freedom give rise to seven families of adsorption conformations (Fig. 1c). We show in the following that the structure of molecular aggregates observed at low coverage as well as the structure of the monolayer (ML) depend not only on *intermolecular* interactions but also in an essential way on these *intramolecular* articulations. Thus, the aim of this work is to gain a better understanding of the successive steps of self-organization of large and jointed organic molecules interacting via van der Waals (vdW) forces. Experimental STM results are compared to calculations based on a semi-classical atomistic model including both molecule-molecule and molecule-substrate interactions.

A comparison between experimental and calculated STM images already allowed to parametrize the molecule/surface interaction and to extract the geometry of the legs while adsorbed on Cu(100). This work pointed out strong deformation of the σ-bonds linkages.[19] Contrary to the case of adsorbed rubrene molecules whereby the intramolecular steric hindrance induces a twisted conformation of central tetracene backbone,[5] the *VL* board stays almost flat above the surface (Fig. 1). With the MM4(2003) force-field,[21] the physisorption energy is found to be around 145 kcal/mol. This fairly high value originates mainly from the attractive vdW board-surface interaction, which induces a rotation of the spacer legs by ± θ as shown in Fig. 1a and 1b, so that the adsorption height of the board is only 3.5 Å instead of the supposed ~ 6 Å of the rigid configuration.[17,19]

Figure 2a shows the surface after deposition of a fraction of monolayer of *VLs*. A single molecule is imaged with four ellipsoidal lobes in a nearly rectangular configuration (see the molecule in the square) corresponding to the DTB groups, whereas the board is not resolved because it is not well enough electronically coupled to the metal substrate.[15,19,22,23] At RT, *VLs* are mobile, thus individual molecules are imaged in a fuzzy way as marked by the arrow. More stable images are obtained when molecules are trapped either by defects like monoatomic steps or in an arrangement with other molecules (Fig. 2a). *VLs* tend to self-organize in *two types* of ordered structures, *entangled or aligned* (full and dotted circles in Fig. 2a). The *entangled state* presents a square configuration involving at least four *VLs*. In this structure, a fluoranthene group of a given molecule points towards the center of the board between the two DTB groups of another, approximately perpendicular molecule (Fig. 2d). It may be noted that the square structure is chiral. As expected from the symmetry of the substrate and the *VL*, both handednesses were observed. In the *aligned state*, the boards are parallel to each other, at 15 degrees from the <110> directions of Cu(100). These two structures are stable up to 20 mn at RT. Squares are assembled when two molecules attach to a former couple of *VLs* in a perpendicular configuration. Such a couple referred to as T-configuration, is much less stable, up to a minute, than the complete square. At low temperature, it seems that this structure was not observed with smaller Landers[16] and is specific of self-assembly of *V*Ls at RT.

Figure 3 shows two terraces separated by a monoatomic step and covered by one ML of *V*Ls, organized in rows up to 50 nm long. In their turn, rows form ordered domains where they are arranged parallel to each other and separated by a distance $a = 3.2 \pm 0.2$ nm. As for the aligned structure mentioned previously (Fig. 2), the rows are oriented at 15 degrees from <110> directions with an intermolecular distance of $1.95 \pm 0.10$ nm. The interaction between two adjacent molecules in a row is not interrupted by the presence of a step, so that domains may spread over two or more terraces. This observation suggests that intermolecular interactions dominate over molecule substrate interactions at ML coverage. The intramolecular contributions generate the self-assembled structure, weakly modulated by the (100) surface, that produces the different domains (Fig. 3a). No square structure can be discerned in the complete monolayer.

Figures 2b and 2c display the structures obtained by molecular mechanics simulations. These snapshots are calculated from a set of starting structures including four molecules, and built in the following way. Each molecule initially belongs to one of the seven families presented in Fig. 1c and stays in this family during relaxation because the simulation does not explore the entire potential energy landscape. The four molecules are placed on the surface in one of the following three initial configurations: (i) four molecules with approximatively perpendicular boards (quasi-entangled structure), (ii) two remote couples with each two approximatively perpendicular boards (two T-configurations), (iii) a row with four approximatively parallel boards (quasi-aligned structure). Then, the intermolecular distances and molecular orientations are allowed to evolve during subsequent energy minimization (see supporting information).

The minimum energy configuration of the square entangled structure is obtained when the four molecules are in the crossed conformation (F family, Fig. 2b and first row of Table 1). This indicates that intermolecular interactions induce intramolecular deformations to locally minimize the energy of the system, leading to this transient structure. Indeed, the pairs of legs inside the square form arches (Fig. 1b) where the distance between the methyl groups in contact with the surface is the greatest. This allows the two perpendicular boards to get closer (Fig. 2d) to maximize attractive interactions with each other and with the upper part of the arch. On the contrary, board-board interactions are hindered if these methyl groups are too close to each other due to steric crowding, as occurs in A, B, D and E conformations. As an illustration, the energy difference between a square with four A molecules and a square with four F molecules is greater than 20 kcal/mol. The distance between legs provides just enough space for another molecule to access, forming this T-configuration (Fig. 2d). Such a key-lock complementarity is not observed with shorter or longer Lander molecules.[16] It is also interesting to note that because of the surface corrugation, the entangled structure is not perfectly symmetric: the boards are not strictly perpendicular, as observed on STM images. Table 1 also shows that such a configuration is privileged over two remote T-couples, as concluded from STM observations.

Dealing now with the row configuration (Fig. 2c, second row of Table 1), the large difference in intermolecular energies between the A and F families may seem surprising. In this case board-board interactions are very weak, and the optimal packing for leg-leg interactions is obtained with parallel legs (A family), which allow a reduced intermolecular distance of 1.85 nm in agreement with experiments (1.95 nm). For the F family, the leg-leg packing between two molecules leads to a larger intermolecular distance (1,93 nm), and thus to a lower intermolecular energy.

These two types of structures (entangled and aligned) act as nucleation seeds for larger clusters. When two *VL*s are added to a square following its particular configuration, a network of squares is built which may extend in two dimensions. *VL*s may also join an aligned couple to form rows up to three elements at this coverage. A mixture of the two cases is illustrated in Fig. 2e: six molecules are arranged to form two squares, one of them acting as a starting point for a row of three molecules.

On purely energetic grounds, we infer from Table 1 that the square configuration is preferred at low coverages. However, we deal here with a single and isolated square. While adding a molecule to an existing row always increases the intermolecular energy by the same amount, adding molecules to form a square adjacent to an existing one is a more complex process. The optimal conformation for *VL*s in a square occurs with inside legs opened (i.e. forming an arch) and outer legs closed (i.e. with opposite angles). It is still possible to build a second square in this case, but with an energy penalty between 1.5 kcal/mol and 4 kcal/mol due to the fact that the two additional molecules have to fit the conformation of the first two ones, with insider legs opened instead of being closed. This second square will be much less stable than the first one. This is the reason why never more than three adjacent squares are observed on STM images. Beyond this explanation, there is also a trivial packing effect which explains the row assembly at ML coverage. At low coverage, the single square configuration minimizes the intermolecular energy per molecule. At high coverage however, the row configuration, with lower per-surface intermolecular energy (higher compactness) is favored. Indeed, the mean area occupied by one molecule is around 8 nm² in the adjacent square configuration versus 5.5 nm² only in the parallel row assembly.

After energy minimization, the equilibrium distance between the terminal hydrogens of two neighbouring molecules belonging to adjacent rows is close to 0.35 nm, which corresponds to the excluded volume around the hydrogen atoms plus a small distance corresponding to the preferred positions of the molecules respective to the copper lattice. The attractive interaction between two molecules in this configuration is indeed very weak (1 kcal/mol). Given the length of a single molecule (2.58 nm), the expected separation between two rows is between 2.9 and 3.0 nm, in agreement with STM observations.

In summary, we have elucidated new mechanisms governing the self-assembly of large, jointed and 3-dimensional molecules on a metallic surface. We have shown that most of the features of self-organization of such molecules can be explained on purely enthalpic grounds using a simple classical force field. Nevertheless, molecules cannot simply be considered as rigid objects, but it is mandatory to take into account their internal molecular degrees of freedom in order to understand the observed structures. Our findings on the transient molecular configurations before a monolayer is completed may have an impact on the design of specific molecules for the controlled self-assembly of new molecular nanostructures. Such a knowledge on the conformation of large adsorbed molecules will give innovative ideas for a molecular design able to build specific mesh on a surface, for instance.

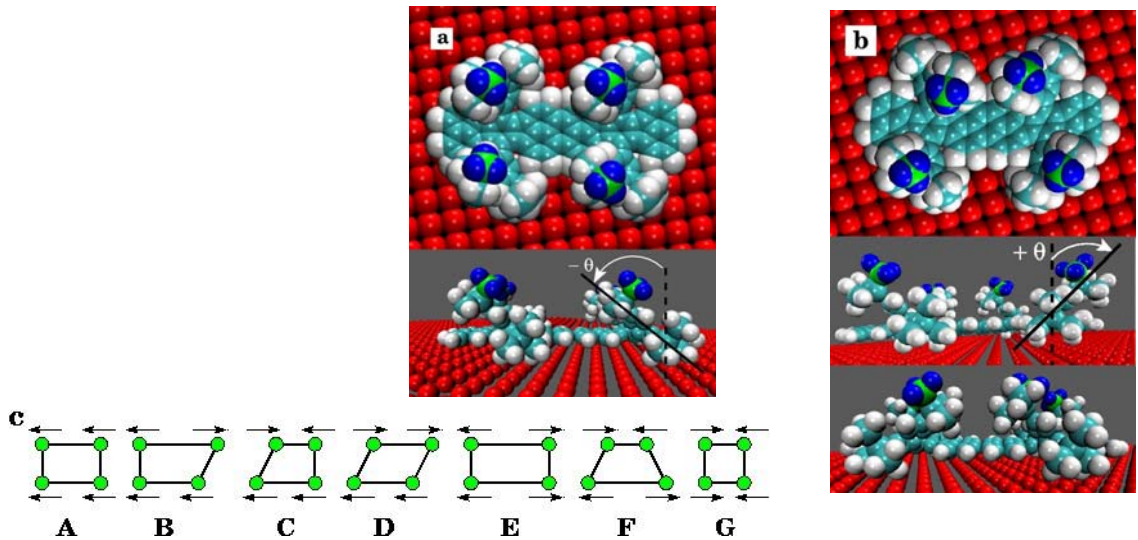

***Figure 1.*** Top and profile views with parallel in a) and crossed in b) configurations of the legs. The colors of the uppermost methyl group were changed for clarity. c) The seven idealized families of adsorbed conformations of the *VL* molecule obtained by rotating each leg by ± θ. Each circle represents one of the three terminal methyl groups of each leg. The arrows point to this group. The calculated conformation in a) belongs to A and in b) to F family, respectively.

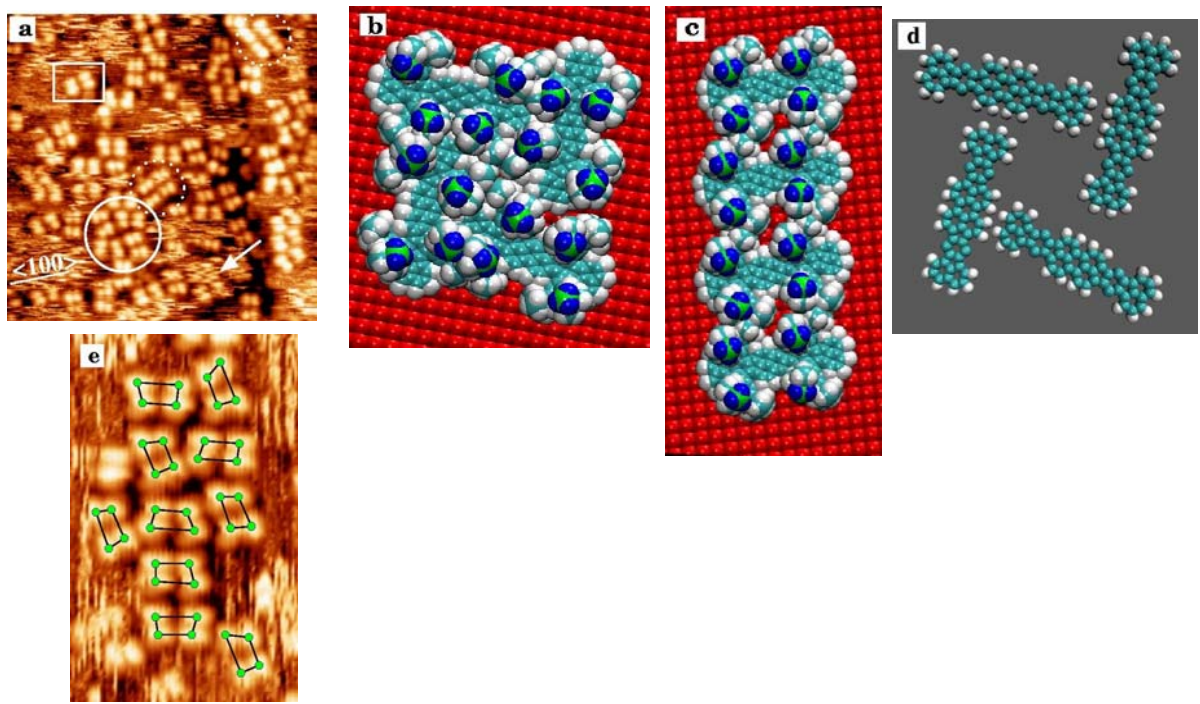

***Figure 2.*** a) STM image (27 nm × 27 nm, $I$ = 10 pA, $U$ = 2.1 V) of *VL*s showing single molecules (inside the white square), entangled structures (full circle), aligned configurations (dotted circles) and diffusing molecules (arrow). b) and c) Simulation snapshots of the entangled and the aligned structures. d) Boards alone from the structure in b). e) STM image (12 nm × 27 nm, $I$ = 10 pA, $U$ = 2.1 V) of a transient molecular aggregate with a mix of structures.

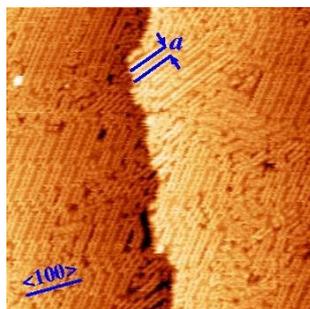

*Figure 3.* STM image (77 nm × 77 nm, *I* = 10 pA, *U* = - 1.5 V) of one ML of *V*ls deposited at RT and annealed up to 100°C.

*Table 1.* Intermolecular energies $E_{LL}$ (in kcal/mol) of structures with four molecules. $E_{LL} = E_{total} - E_{SL} - E_{IL}$, where subscripts SL and IL refer to *Surface-Lander* and *Intra-Lander* vdW interactions respectively.

| Configuration | Parallel legs | Crossed legs |
|---|---|---|
| Square | - 31 | - 55 |
| Row | - 37 | - 12 |
| Two T-structures | - 28 | - 46 |

*Supplementary materials*

The Cu(100) sample was cleaned according to the usual cycling procedure of $Ar^+$ ion sputtering (600 eV, 4 μA $cm^{-2}$) followed by annealing at 500 °C. Molecules were sublimated from a tantalum crucible heated with a direct current. During the sublimation, the sample was at RT and eventually it was heated afterwards. STM experiments were carried out at RT.

The calculations presented in this paper rely on a model consisting of a fixed Cu(100) surface, with copper atoms lying on their ideal lattice position, and one to four *VL* molecules deposited on this surface in various initial configurations. Only the two uppermost copper layers are included in the computations. The lateral dimension of the simulation cells, including at least 4944 atoms, is 11.6 nm. The non-bonded interactions are truncated to a radius of 0.95 nm, and two-dimensional periodic boundary conditions are applied. The limited memory $BFGS^1$ method was employed for energy minimization. The initial configurations, deemed to relaxation on the surface, cover the squares, rows and T-clusters observed experimentally. The molecules are placed close enough to each other and to their guess structure (aligned or entangled), so that the gradient of attractive or repulsive forces is large enough to allow starting and convergence of the energy minimization. For example, in the case of aligned molecules, an initial separation distance of 1.4 or 2.3 nm will result in the same equilibrium separation of 1.85 nm. The individual *VL* molecules initially belong to one of the seven different families of conformations on the surface (Fig. 1). All these structures are relaxed on the surface using the DL_POLY software package.[2] During the relaxation, the position and orientation of the molecules relative to the surface will evolve, but in most cases no significant change of individual *VL* conformation is noticed, that is the molecule stays in the same conformation family. The forcefield employed here is based on $MM4(2003)^3$ for intermolecular and intramolecular interactions between *VL* atoms. The rotation barrier for a *VL* molecule over the surface is computed over a range of 180 degrees, by placing one molecule over the surface at evenly spaced initial guess orientations separated by 10 degrees. These eighteen structures are first relaxed over the surface, then rotated stepwise by one degree, each step being followed by subsequent relaxation. When a stable minimum is reached, the angle between the board axis and the copper compact direction converges to the same value whatever the initial orientation.

---